\documentclass[aps,prd,preprint,tightenlines,superscriptaddress,
amsfonts,amssymb,amsmath,byrevtex,showpacs,nofootinbib]{revtex4-1}
\usepackage{amssymb,amsmath}
\usepackage{latexsym}
\usepackage{graphicx}
\usepackage{geometry}
\usepackage{epstopdf}
\usepackage{color}
\usepackage{subcaption}

\begin{document}
\newcommand{\dN}{\mathbb N}
\newcommand{\dR}{\mathbb R}
\newcommand{\dC}{\mathbb C}
\newcommand{\dS}{\mathbb S}
\newcommand{\dZ}{\mathbb Z}
\newcommand{\id}{\mathbb I}
\newcommand{\bea}{\begin{eqnarray}}
\newcommand{\eea}{\end{eqnarray}}

\title{Comparing the dynamics of diagonal and general \\ Bianchi IX spacetime}

\author{Ewa Czuchry}
\email{ewa.czuchry@ncbj.gov.pl}\affiliation{ Department of Fundamental
Research, National Centre for Nuclear Research, Ho{\.z}a 69,
00-681 Warsaw, Poland}

\author{Nick Kwidzinski}
\email{nk@thp.uni-koeln.de}\affiliation{Institut f\"{u}r Theoretische Physik,
Universit\"{a}t zu K\"{o}ln, Z\"{u}lpicher Strase 77, 50937 K\"{o}ln, Germany}

\author{W{\l}odzimierz Piechocki}
\email{wlodzimierz.piechocki@ncbj.gov.pl}\affiliation{ Department
of Fundamental Research, National Centre for Nuclear Research,
Ho{\.z}a 69, 00-681 Warsaw, Poland}

\date{\today}

\begin{abstract}
We make comparison of the dynamics of the diagonal and nondiagonal
Bianchi IX models in the evolution towards the cosmological singularity.
Apart from the original variables, we use the Hubble normalized ones commonly
applied in the examination of the dynamics of homogeneous models.
Applying the dynamical systems method leads to the result that in both
cases the continuous space of critical points is higher dimensional and
they are of the nonhyperbolic type. This is a generic feature of
the dynamics of both cases and seems to be independent on the choice
of phase space variables. The topologies of the corresponding
critical spaces are quite different. We conjecture that the nondiagonal
case may carry a new type of chaos different from the one specific
to the usually examined diagonal one.
\end{abstract}


\maketitle


\section{Introduction}

According to the singularity theorems of General Relativity (GR), the   evolution of an expanding universe is
geodesically {past}-incomplete. The Belinskii, Khalatnikov and Lifshitz (BKL) \cite{BKL22,BKL33} scenario predicts
that on approach to a  space-like {cosmological} singularity  the dynamics of gravitaional field simplifies
as time derivatives in Einstein equations dominate over  spatial derivatives (see \cite{Gar} for numerical support
for BKL). In this regime the evolution of the Universe becomes strongly non-linear and chaotic, comprising
expanding and contracting oscillatory phases  around the singular point. One believes that an imposition
of quantum rules onto this scenario may heal the singularity. Finding the  nonsingular quantum BKL scenario would
mean solving, to some extent,  the generic cosmological singularity problem. Such a quantum theory could be used
as a realistic model of the very early Universe.

Quantization of the BKL scenario should be preceded by the quantization of the Bianchi IX model. This seems to be a
reasonable strategy because the BKL scenario has been obtained via analysis of the dynamics of the Bianchi IX spacetime.
The  three metric on space of the Bianchi IX model (in the synchronous reference system) is in general nondiagonal
for general matter models. However in the case of vacuum or simple fluids it can be diagonalized during the entire evolution of
the system. We refer to these two cases as nondiagonal and diagonal Bianchi IX models, respectively.
The best prototype for the BKL scenario is the   nondiagonal Bianchi IX model \cite{BKL33,Belinski:2014kba,BKL2}
corresponding to general matter fields.

The quantization of the Bianchi IX model requires full understanding of its classical dynamics in terms of variables
convenient for quantization procedure. Our recent paper \cite{Czuchry:2012ad} has initiated such analysis. As far as we
\cite{Bogo1,Heinzle:2009du} and references therein). The examination of the dynamics presented in \cite{Bogo2} of the
nondiagonal case is mathematically satisfactory, but seems to be too complicated to be used in  any quantization scheme.

Recent analysis indicate that the dynamics of the nondiagonal case has asymptotic regime near the singularity \cite{Nick}.
The dynamics of this regime looks similarly to the dynamics of the diagonal case (devoid of asymptotic regime).
However, the symmetry aspects of both set of equations defining the corresponding dynamics are quite different,
which leads to the different topologies of the corresponding spaces of solutions. The aim of this paper is the examination
of these differences in more details.

In this paper we use two quite different sets of variables
parameterizing the dynamics:  original BKL type \cite{Belinski:2014kba,BKL2}
and quasi Hubble normalized \cite{Heinzle:2009eh}. Making use  the scale
invariance of Einstein equations one can introduce variables which divided
by the Hubble parameter become scale invariant \cite{Uggla:2003fp}. The Hubble
parameter, which in general spacetime is a geometrical average of expansion
rates in three space directions, becomes infinite approaching the singularity.
Although gravitational field variables like orthonormal frame variables also
diverge approaching singularity, normalized by Hubble parameter remain finite
and more useful for  analytical analysis \cite{Heinzle:2009du,Uggla:2003fp},
and they enabled successful numerical verification \cite{Gar}.

However, original BKL variables and Hubble normalized ones cannot be connected
by canonical transformation. In both cases, applying dynamical systems method
enables  identification of the spaces of non-isolated critical (equilibrium)
points, which are of  nonhyperbolic type.  Topologies of these
spaces are  quite different, and making them explicit constitutes one of
the main results of this paper. Additional result is expressing the asymptotic
nondiagonal Bianchi IX model in terms of non-divergent variables similar to the
Hubble normalized variables, thus enabling future more detailed investigations.

Our paper is organized as follows: Section II concerns the
nondiagonal case. We introduce quasi Hubble normalized variables,
examine the asymptotic dynamics in these (and BKL) variables, and identify
the spaces of critical points of the corresponding vector fields.
The diagonal case is considered in Sec. III, where we follow  the
steps of Sec. II. The numerical simulations of the dynamics is presented in Sec. IV.
We conclude in Sec. V.
Appendix A concerns the issue of an effective form of the metric
near the singularity. The choice of quotient coordinates,
presented in App. B, enables making an extension of the
interpretation of our results.  We present the relationship
between the BKL and  our new variables in App. C. Finally,  we
apply the Poincar\'{e} sphere to deal with the space of critical
points in finite region of phase space in App. D.

\section{The nondiagonal case}

The general form of a line element of the { nondiagonal}
Bianchi IX model, in the synchronous reference system, reads
\begin{equation}\label{dd1}
ds^2 = dt^2 - \gamma_{ab}(t)e^a_\alpha e^b_\beta dx^\alpha
dx^\beta ,
\end{equation}
where Latin indices $a,b,\ldots$ run from $1$ to $3$ and label the
frame vectors $e^a_\alpha$, and Greek indices
$\alpha,\beta,\ldots$ take values $1,2,3$ and concern space
coordinates, and where $\gamma_{ab}$ is a spatial metric.

It was shown in \cite{BKL22,BKL33} that near the cosmological
singularity the { general} form of the metric $\gamma_{ab}$
should be considered. Consequently, one cannot globally
diagonalize the metric, i.e.  for all values of time.  After
making use of the Bianchi identities, freedom in the rotation of
the metric $\gamma_{ab}$ and frame vectors $e^a_\alpha$, one
arrives at the well-defined, but complicated system of equations
specifying the dynamics of the  nondiagonal Bianchi IX model
\cite{BKL2}.  Assuming that the {anisotropy} of space
may grow without bound,  when approaching the singularity, enables
considerable simplification of the dynamics. Finally, the {
asymptotic} form (near the cosmological singularity) of the
dynamical equations of the { nondiagonal} Bianchi IX model
reads \cite{Belinski:2014kba,BKL2,Czuchry:2012ad}:

\begin{equation}\label{L1} \frac{d^2 \ln a }{d
\tau^2} = \frac{b}{a}- a^2,~~~~\frac{d^2 \ln b }{d \tau^2} = a^2 -
\frac{b}{a} + \frac{c}{b},~~~~\frac{d^2 \ln c }{d \tau^2} = a^2 -
\frac{c}{b},
\end{equation}
where $a,b,c$ are functions of time $\tau$, satisfying  the
constraint
\begin{equation}\label{L2}
\frac{d \ln a}{d \tau}\;\frac{d \ln b}{d \tau} + \frac{d \ln a}{d
\tau}\;\frac{d \ln c}{d \tau} + \frac{d \ln b}{d \tau}\;\frac{d
\ln c}{d \tau} = a^2 + \frac{b}{a} + \frac{c}{b},
\end{equation}
and where $\tau$ is connected with the cosmological time variable
$t$ as follows
\begin{equation}\label{dd2}
    dt = \sqrt{\gamma}\; d\tau
\end{equation}
($\gamma$ denotes  the determinant of $\gamma_{ab}$).

Turning the above dynamics  into Hamiltonian dynamics, one can
examine qualitatively the mathematical structure of the
corresponding physical phase space by using the dynamical systems
method (DSM). It has been found  that  the {\it critical} points
of the system have the following properties: (i) define a
three-dimensional  { continuous}  subspace of $\bar{\dR}^6$
defined by the relation $a \gg b \gg c > 0$, with $a \rightarrow
0$ (see, Eq. (38) of \cite{Czuchry:2012ad} for more details), and
(ii) are of the { nonhyperbolic} type.

The property (i) was already found long time ago \cite{BKL2}
without using the DSM. The characteristic (ii) has been identified
recently \cite{Czuchry:2012ad}. The latter property means that
getting insight into the structure of the space of orbits near
such critical set requires further examination of the exact {
nonlinear} dynamics.  So the results obtained from
{inearization} of the dynamics cannot be conclusive (see, e.g.,
\cite{Wiggins}).

\subsection{Quasi Hubble normalized variables}

To make progress in understanding the structure  of our
critical set, we propose the parametrization of the dynamics by an
analog of the so-called Hubble-normalized (HN) variables $(\tilde{\Sigma}_\alpha,\tilde{N}_\alpha)$ (see,
e.g., \cite{Heinzle:2009eh}-\cite{WE} and references therein).
They can be ascribed  to  the {vacuum} Bianchi type
models in which case the spatial metric can be taken to be {
diagonal}.  Assuming a spacetime admitting a foliation $\mathcal{M}\mapsto
\Sigma\times\mathbb{R}$, where $\Sigma$ is spacelike, the line element of the spatially
homogenous Bianchi type model reads, following the original notation of
\cite{Heinzle:2009eh}-\cite{WE}: \begin{equation}\label{eq}
d s^2= - d t^2+
g_{11}(t)\ {\omega^1}\otimes{\omega^1}+g_{22}(t)\ {\omega^2}\otimes{\omega^2}+g_{33}(t)\ {\omega^3}\otimes{\omega^3}\, ,
\end{equation}
where the $\omega^\alpha$'s are 1-forms on $\Sigma$ invariant with
respect to the action of a simply transitive group of motions on
the leaf and subject to
\begin{equation}\label{cartan}
d\omega^1=-\hat{n}_{1}\ \omega^2\wedge\omega^3\, ,
d\omega^2=-\hat{n}_{2}\ \omega^3\wedge\omega^1\, ,
d\omega^1=-\hat{n}_{3}\ \omega^1\wedge\omega^2\,
\end{equation}
where $\hat{n}_{\alpha}$ are structure constants of the corresponding Lie
algebra. In case of the Bianchi IX model
$\hat{n}_{\alpha}=1$.

Within this framework one can  define the expansion $\theta$ and the shear $\sigma^\alpha_{\ \beta}$:
\begin{equation}\label{L8}
\theta:=-\textrm{tr} (k)\, ,\ \ \sigma^\alpha_{\ \beta}:=
-k^\alpha_{~\beta} +\frac{1}{3}\,
\textrm{tr}    (k) \,\delta^\alpha_{~\beta} = \textrm{diag}(\sigma_1, \sigma_2.
    \sigma_3),
\end{equation}
where $k_{\alpha\beta}$ is the second fundamental form associated with hypersurfaces $\{t=\textrm{const.}\}$ and $\sigma_\alpha$ fulfill $\Sigma_\alpha\sigma_\alpha=0$.
The Hubble variable $H$ is proportional to the expansion $ H= \theta/3$ and is related to changes of the spatial volume density via
$d \sqrt{g}/dt= 3 H \sqrt{g}$, where $g=\textrm{det} g_{\alpha\beta}$. One can also define variables $n_\alpha$
\begin{equation}
n_\alpha=
\hat{n}_\beta
\frac{g_{\alpha}^{\ \beta}}{\sqrt{g}}.
\end{equation}
For the Bianchi IX model  there exists a one-to-one correspondence between the set of the definded above variables $(H, \sigma_\alpha, n_\alpha)$ (with $\Sigma_\alpha\sigma_\alpha=0$)  and the standard ones $(g_{\alpha\beta},k_{\alpha\beta})$.
In this setting one can introduce the Hubble normalized (HN) variables $(\tilde{\Sigma}_\alpha,\tilde{N}_\alpha)$ (here we use symbol $\tilde{}$ for distinguishing the original variables and our subsequent ones),  which are orthonormal frame variables
$\sigma_\alpha$ and $n_\alpha$ normalized by the Hubble variable $H$:
\begin{equation}\label{HN}
\tilde{\Sigma}_\alpha:= \frac{\sigma_\alpha}{H}, \ \ \tilde{N}_\alpha :=\frac{n_\alpha}{H},
\end{equation}
These are dimensionless quantities which fully describe the dynamics of the three-dimensional spacelike hypersurface $\Sigma$. Near the singularity, where space curvature and expansion all diverge, the HN  variables remain finite, as dividing by divergent Hubble variable $H$ factors out the overall expansion. Analysing dynamics of the Bianchi IX spacetimes near its singularity in terms of HN variables brought a lot of important and interesting results (see, e.g., \cite{Heinzle:2009du,Heinzle:2009eh} and references therein).

Henceforth, it would be natural trying to formulate dynamics of the non-diagonal Bianchi IX model in terms of the HN variables. However there is the key difficulty laying in the definition of those variables, formulated for diagonal metrics, in case of the general Bianchi IX spacetime described by the metrics  (\ref{dd1}). This metrics is generally non-diagonal globally, although it can be diagonalized at each separate moment of time.  According to \cite{BKL2} the exact 3-dimensional metric $\hat{\gamma}$ is
given by
\begin{equation}\label{h1}
\hat{\gamma} = \hat{R}^{-1}\hat{\Gamma}\hat{R},
\end{equation}
where $\hat{\Gamma} = \textrm{diag} (\Gamma_1, \Gamma_2, \Gamma_3)$ and
$\hat{R}$  is an orthogonal
matrix ($\hat{R}^T = \hat{R}^{-1},\; \det \hat{R} = 1$). The matrix
$\hat{R}$ transforms the 3-dimensional metric tensor $g_{\alpha \beta}$ to the
principal axes and this rotation might be described in terms of Euler angles $(\theta,
\varphi, \psi)$: rotation, precession and pure rotation. In other words $\hat{R}=
\hat{R}_\theta\hat{R}_\varphi\hat{R}_\psi$, where  $\hat{R}_\theta$, $\hat{R}_\varphi$ and $\hat{R}_\psi$
are standard rotation matrices.

In the general case, the Euler angles $(\theta,
\varphi, \psi)$ are time dependent and describe the rotation with
respect to the frame vectors $e^a$, which are fixed. In the
asymptotic regime the Euler
angles become time independent, but $\Gamma_\alpha$ stay being
functions of time.

One can diagonalize  the metric $\hat{\gamma}$ in the asymptotic
regime by using $\hat{R}\hat{\gamma}\hat{R}^{-1} = \hat{\Gamma}$.
Since $\hat{R}$ is time independent there,  this diagonal form
will exist until the gravitational system approaches the
singularity. In this regime, the line element (\ref{dd1}) can be
presented as follows (see \cite{BKL22,BKL33} for more details)

\begin{equation}\label{h3}
ds^2 = dt^2 - \big( a^2 e_\alpha ^{(1)}e_\beta^{(1)} + b^2
e_\alpha ^{(2)}e_\beta^{(2)} + c^2 e_\alpha ^{(3)}e_\beta^{(3)}
\big) dx^\alpha dx^\beta ,
\end{equation}
where
\begin{equation}\label{h4}
a :=\Gamma_1,~~b:= \Gamma_2 C^2 \cos^2 \theta_0,~~c := \Gamma_3
C^4 \sin^2\theta_0 \cos^2\theta_0 \sin^2\psi_0 ,
\end{equation}
and where $C$ is a constant of motion. The metric (\ref{h3}) describes only
the oscillatory modes devoid of the rotation. Since $a, b$ and $
c$ satisfy Eqs. (\ref{L1}) - (\ref{L2}), derived from the {\it
exact} system of equations with nondiagonal form of 3-metric, they have encoded
nondiagonal aspects of the metric, and the line element:
\begin{equation}\label{L3}
g_{11} := a^2,~~~g_{22}:= b^2,~~~g_{33}:= c^2,~~~g_{\alpha \beta}
:= 0~~~\mbox{if $\alpha \neq \beta$},
\end{equation}
may be
interpreted as presenting an {\it effective}  3-metric.
This identification suggests that we have  a sort of an {
effective} diagonal metric $g_{\alpha \beta}$ near the
cosmological singularity, i.e., in the { asymptotic} region of
spacetime.

The effective 3-metric (\ref{L3}) is used below to
introduce quasi-HN (qHN) variables.
In this settings we define the new variables
$(N_\alpha,\Sigma_\alpha)$ as follows:
\begin{equation}\label{a1}
    N_1 := a^2 V,~~~N_2 := b^2 V,~~~N_3 := c^2 V,
\end{equation}
\begin{equation}\label{a2}
\Sigma_1 := V \frac{d}{d \tau} \ln a  -1,~~~\Sigma_2 :=
V\frac{d}{d \tau} \ln b - 1,~~~\Sigma_3 := V \frac{d}{d \tau}\ln c
- 1 ,
\end{equation}
where $V = 3/\frac{d}{d \tau}\ln (a b c)$, and where $(a,b,c)$
satisfy Eqs. (\ref{L1}) and (\ref{L2}). Thus, $\Sigma_1 + \Sigma_2
+\Sigma_3 = 0$ identically, and $N_1 > 0, N_2 > 0, N_3 > 0$ as $a
b c  \rightarrow 0$ near the singularity.

In what follows we will present similarities between the set of defined above variables and original HN ones.

The second fundamental form $k_{\alpha \beta}$ associated with
(\ref{L3}) is defined to be
\begin{equation}\label{L6}
k_{\alpha \beta}:= - \frac{1}{2}\frac{d}{dt} g_{\alpha \beta} = -
\frac{1}{2} \frac{1}{abc}\frac{d}{d\tau} g_{\alpha \beta} =:-
\frac{1}{2 v} \dot{g}_{\alpha \beta},
\end{equation}
where due to (\ref{dd2}) we have
\begin{equation}\label{L5}
    dt/d\tau := \sqrt{g} = a b c =: v,
\end{equation}
and where $g:= \det g_{\alpha \beta}$, so $v$ is the spatial
volume density.

If we take $k^\alpha_{~\beta} := g^{\alpha \gamma} k_{\gamma
\beta}$, the trace of the matrix $k_{\alpha \beta}$ reads
\begin{equation}\label{L7}
    \textrm{tr} (k) = k^\alpha_{~\alpha}=
    -\frac{1}{abc}\Big(\frac{\dot{a}}{a}+\frac{\dot{b}}{b}+\frac{\dot{c}}{c}\Big)
    = - \frac{1}{v}\frac{d}{d\tau} \ln v.
\end{equation}

Defining  the expansion $\theta$ by
\begin{equation}\label{L9}
\theta:= \frac{d}{dt} \ln \sqrt g = \frac{1}{v}\frac{d}{d\tau}
 \ln v,
\end{equation}
we get $\theta := - \textrm{tr} (k)$. The volume changes according to $d
v/dt = \theta\, v $. Following the considerations in
\cite{Heinzle:2009eh,Geo}, we define the Hubble variable
\begin{equation}
 H:= \frac{\theta}{3}=\frac{1}{3v}\frac{d}{d\tau}
\ln v.
\end{equation}

Thus the variables defined
in Eqs. (\ref{a1}) and (\ref{a2}) coincide with Hubble normalized variables, namely:
\begin{equation}\label{L10}
\Sigma_\alpha := \frac{\sigma_\alpha}{  H} = \frac{\sigma^{\bar{\alpha}}_{~\bar{\alpha}}} { H}:=
\frac{\big(- k^{\bar{\alpha}}_{~\bar{\alpha}} +\frac{1}{3}\, \textrm{tr} (k) \big)} { H},
\end{equation}
where bared indices denote no summation convention, and  $\Sigma_1 + \Sigma_2 + \Sigma_3 = 0$ identically. We also have
\begin{equation}\label{L11}
N_\alpha :=\frac{ n_\alpha }{H},~~~\mbox{where}~~~~n_\alpha := \frac{g_{\bar{\alpha}
\bar{\alpha}}}{ \sqrt{g}} .
,
\end{equation}
directly corresponding to the definition  (\ref{HN}) in our effective 3-metrics.

\subsection{Dynamics}

\subsubsection{Finding the vector field}

In what follows we derive  the vector field corresponding to
(\ref{L1}) - (\ref{L2}) entirely in terms of the qHN variables.
Acting with $d/d\tau$ on  (\ref{a1}) and making use of (\ref{L1})
leads to the following set of equations
\begin{equation}\label{nn3}
\dot{N}_\alpha =  N_\alpha (2\pi_\alpha - \frac{1}{3}N_1),
~~~\alpha = 1,2,3.
\end{equation}
One can rewrite (\ref{a2}) as follows
\begin{equation}\label{nc}
\Sigma_\alpha +1 =  3 \pi_\alpha/f,~~~\alpha = 1,2,3,
\end{equation}
where $f := \pi_1 + \pi_2 + \pi_3$. Inserting  (\ref{nc}) into
(\ref{nn3}) yields
\begin{equation}\label{J1}
    \dot{N}_\alpha = \frac{N_\alpha}{3}\big(2 (\Sigma_\alpha +1)f - N_1
    \big).
\end{equation}

Acting with $d/d\tau$ on both sides of (\ref{nc}) and using
(\ref{L1}) gives
\begin{eqnarray}
  \label{an1} \dot{\Sigma}_1 &=& - \frac{N_1}{3}(4 + \Sigma_1) +
  \frac{3}{f}\sqrt{\frac{N_2}{N_1}},\\
  \label{an2} \dot{\Sigma}_2 &=&  \frac{N_1}{3}(2 - \Sigma_2) +
  \frac{3}{f}\big( \sqrt{\frac{N_3}{N_2}}-\sqrt{\frac{N_2}{N_1}}\big),\\
  \label{an3} \dot{\Sigma}_3 &=&  \frac{N_1}{3}(2 - \Sigma_3) -
  \frac{3}{f}\sqrt{\frac{N_3}{N_2}}.
\end{eqnarray}
Due to (\ref{sol1}), and $\Sigma_1 + \Sigma_2 + \Sigma_3 =0$, we have
\begin{equation}\label{J2}
f  = 3 \Pi .
\end{equation}
Inserting (\ref{J2}) into (\ref{J1}) - (\ref{an3}), we finally
obtain the following vector field specifying the dynamics entirely
in the qHN variables:
\begin{eqnarray}
  \label{JH1} \dot{N}_1 &=& 2 \Pi N_1 \;(1+\Sigma_1)  - \frac{N_1^2}{3},\\
  \label{JH2} \dot{N}_2 &=& 2 \Pi N_2 \;(1+\Sigma_2)
  - \frac{N_1 N_2}{3}, \\
  \label{JH3} \dot{N}_3 &=&  2 \Pi N_3{(1+\Sigma_3)} - \frac{N_1 N_3}{3},\\
  \label{JH4} \dot{\Sigma}_1 &  = &  \frac{N_1}{3}(-4 -\Sigma_1)
  + \frac{1}{\Pi}\sqrt{\frac{N_2}{N_1}},\\
\label{JH5} \dot{\Sigma}_2 &  = &  \frac{N_1}{3}(2 - \Sigma_2)
  + \frac{1}{\Pi}\Big(\sqrt{\frac{N_3}{N_2}} -
  \sqrt{\frac{N_2}{N_1}}\Big),\\
\label{JH6} \dot{\Sigma}_3 &  = & \frac{N_1}{3}(2 - \Sigma_3)
  - \frac{1}{\Pi}\sqrt{\frac{N_3}{N_2}},
\end{eqnarray}
where $\Sigma_1 + \Sigma_2 + \Sigma_3 = 0$.  The variable $\Pi$
has to satisfy the constraint (\ref{conPi}), which corresponds to
the original constraint (\ref{L2}). Taking into account the
constraint yields the system of equations:

\begin{eqnarray}
\dot{N}_1&  = &-\frac{{N_1}^2}{3}-\frac{N_1(1+{\Sigma_1})  \left({N_1}+\sqrt{{N_1}^2-4
\Sigma
\left(\sqrt{\frac{{N_2}}{{N_1}}}+\sqrt{\frac{{N_3}}{{N_2}}}\right)}\right)}{\Sigma},\label{fin1}\\
\dot{N}_2 &  = &-\frac{{N_1} {N_2}}{3}-\frac{N_2(1+{\Sigma_2})  \left({N_1}+
\sqrt{{N_1}^2-4 \Sigma
\left(\sqrt{\frac{{N_2}}{{N_1}}}+\sqrt{\frac{{N_3}}{{N_2}}}\right)}\right)}{\Sigma},\label{fin2}\\
\dot{N}_3&  = &-\frac{{N_1} {N_3}}{3}-\frac{N_3(1-{\Sigma_1}-{\Sigma_2}) \left({N_1}+
\sqrt{{N_1}^2-4 \Sigma
\left(\sqrt{\frac{{N_2}}{{N_1}}}+\sqrt{\frac{{N_3}}{{N_2}}}\right)}\right)}{\Sigma},\label{fin3}\\
\dot{\Sigma}_1&  = &\frac{1}{3} (-4-{\Sigma_1}) {N_1}-\frac{2 \Sigma\sqrt{\frac{{N_2}}{{N_1}}}}{{N_1}+\sqrt{{N_1}^2-4
\Sigma \left(\sqrt{\frac{{N_2}}{{N_1}}}+
\sqrt{\frac{N_3}{N_2}}\right)}},\label{fin4}\\
\dot{\Sigma}_2&  = &\frac{1}{3} (2-{\Sigma_2}) {N_1}+\frac{2
\Sigma
\left(\sqrt{\frac{{N_2}}{{N_1}}}-\sqrt{\frac{{N_3}}{{N_2}}}\right)}{{N_1}+\sqrt{{N_1}^2-4
\Sigma
\left(\sqrt{\frac{{N_2}}{{N_1}}}+\sqrt{\frac{{N_3}}{{N_2}}}\right)}}\label{fin5}
\end{eqnarray}
where $\Sigma:=-3+{\Sigma_1}^2+{\Sigma_1} {\Sigma_2}+{\Sigma_2}^2$.

\subsubsection{Critical points of the dynamics}

Direct inspection of the system (\ref{fin1}) - (\ref{fin5}) leads to the following
identification of the set of
the critical points:

\begin{equation} \label{critnew}S_{qHN}: =\{(
\Sigma_1,\Sigma_2,N_1,N_2,N_3) ~|~ ( N_1 \rightarrow
0, N_2 \rightarrow 0, N_3 \rightarrow 0)\} \subset {\bar{\dR}}^6
,\end{equation} in such a way that $N_3 << N_2 << N_1$ and $\sqrt{N_2/N_1}
<< N_1^2 \rightarrow 0$, and $\sqrt{N_3/N_2} << \sqrt{N_2/N_1}
\rightarrow 0$, which imply that
\begin{equation}\label{cp1}
\sqrt{N_3/N_2} << \sqrt{N_2/N_1} << N_1^2 \rightarrow 0.
\end{equation}

One can avoid taking the uncommon form of the limits (\ref{cp1})
by introducing new variables, which we consider in App. B.
However, this does not change the character of critical points.
They stay to be the nonhyperbolic ones.
A critical point is called a hyperbolic fixed point if  all
the eigenvalues of the Jacobian matrix of the linearized equations
at this point have nonzero real parts. Otherwise, it is called a
nonhyperbolic fixed point \cite{Wiggins}. In the sequence we analyze
the Jacobian for the above system and determine character of critical points.

\subsubsection{The linearization of the vector field}

One may verify, with some effort, that some elements  of the
Jacobian $J$ of the system  (\ref{fin1}) - (\ref{fin5}), evaluated
at any point of $S_{qHN}$, are diverging. This behavior comes from
differentiating  square roots. However, when calculating
characteristic polynomial of the Jacobian $J$ at any point those
divergencies cancel out due to relations \eqref{cp1} giving
\begin{equation}\label{jac1}
P(\lambda)=-\lambda^5,
\end{equation}
so the  eigenvalues are $\left(0,0,0,0,0\right)$. Owing to very complicated form of the Jacobian
matrix $J$ and characteristic polynomial, we exhibit only the result after embedding conditions \eqref{cp1}.
Since the real parts of all eigenvalues of the Jacobian are equal
to zero, we are dealing with the {\it nonhyperbolic} critical
points.

Our system  evolves  asymptotically, as time goes to zero (when
the system approaches the cosmological singularity), to the
nonhyperbolic critical subspace with the coordinates
$(\Sigma_1,\Sigma_2,N_1,N_2,N_3)$ given by
\begin{equation}\label{ncp}
(\Sigma_1,\Sigma_2,0,0,0).
\end{equation}
Further analysis should be based on making use of the {\it exact}
form of our vector field.

\section{The diagonal case}

In what follows we demonstrate that the asymptotic forms of the
dynamics of the non-diagonal and diagonal Bianchi IX model are
quite different.

The dynamics of the diagonal Bianchi IX in asymptotic  regime near the singularity may be obtained from the asymptotic
dynamics of non-diagonal model with zero rotation of principal values $\Gamma_a$ of the three-dimensional metric tensor
$\gamma_{ab}$ around frame vectors $e^a$.
It means that the Euler angles $(\theta,\varphi,\psi)$, describing the rotation with
respect to the frame vectors are fixed
\begin{equation}\label{AA}
    (\theta,\varphi,\psi) = (\theta_0,\varphi_0,\psi_0),
\end{equation}
so they are no longer the degrees of freedom of the system.
In that case the Einstein equations for the general Bianchi IX model in the vicinity of singularity derived in \cite{BKL2} take the following form:
\begin{eqnarray}
 \label{bb1} (\ln \Gamma_1)^{\cdot \cdot} + \Gamma_1^2 - (\Gamma_2 - \Gamma_3)^2&=& 0,\\
 \label{bb2} (\ln \Gamma_2)^{\cdot \cdot} + \Gamma_2^2 - (\Gamma_1 - \Gamma_3)^2  &=& 0,\\
 \label{bb3} (\ln \Gamma_3)^{\cdot \cdot} + \Gamma_3^2 - (\Gamma_1 - \Gamma_2)^2  &=& 0,
\end{eqnarray}
where we assumed that the total angular momentum of the system vanishes, unlike in the general case, with rotation frozen near the singularity but with non-zero total angular momentum.  The constraint equation, coming from the Bianchi identities, reads
\begin{align}\label{bb4}
((\ln \Gamma_1)^{\cdot})^ 2& + ((\ln \Gamma_2)^{\cdot})^ 2 + ((\ln
\Gamma_3)^{\cdot})^ 2 - ((\ln \Gamma_1 \Gamma_2
\Gamma_3)^{\cdot})^ 2\nonumber\\& + 2(\Gamma_1^2 + \Gamma_2^2 + \Gamma_3^2 ) -
4 (\Gamma_1 \Gamma_2 + \Gamma_1 \Gamma_3 + \Gamma_2 \Gamma_3) = 0.
\end{align}
For the comparison with (\ref{L1}) - (\ref{L2}), we rewrite
(\ref{bb1}) - (\ref{bb4}) using  the notation: $\tilde{a}:=
\Gamma_1,~\tilde{b}:= \Gamma_2, \tilde{c}:= \Gamma_3$~, and get
\begin{eqnarray}
 \label{s1} (\ln \tilde{a})^{\cdot \cdot}& = &  (\tilde{b} - \tilde{c})^2  -\tilde{a}^2 ,\\
 \label{s2} (\ln \tilde{b})^{\cdot \cdot}& = & ( \tilde{c} -\tilde{a})^2 -\tilde{b}^2 ,\\
 \label{s3} (\ln \tilde{c})^{\cdot \cdot}& = &  (\tilde{a} - \tilde{b})^2 -\tilde{c}^2 ,
\end{eqnarray}
with the dynamical constraint:
\begin{equation}\label{s4}
((\ln \tilde{a})^{\cdot})^ 2 + ((\ln \tilde{b})^{\cdot})^ 2 +
((\ln \tilde{c})^{\cdot})^ 2 - ((\ln \tilde{a}\tilde{b}
\tilde{c})^{\cdot})^ 2 + 2(\tilde{a}^2 + \tilde{b}^2 + \tilde{c}^2
) - 4 (\tilde{a} \tilde{b} + \tilde{a} \tilde{c} + \tilde{b}
\tilde{c} ) =0.
\end{equation}

The dynamics of the diagonal and
nondiagonal cases are quite different. Let us indicate just one
aspect of this non-equivalence. Namely, it is clear that Eqs.
(\ref{s1}) - (\ref{s4}) are symmetric with respect to the
permutations:
\begin{equation}\label{CC1}
    (\tilde{a},\tilde{b},\tilde{c}) \rightarrow (\tilde{b},\tilde{c},\tilde{a})
    \rightarrow (\tilde{c},\tilde{a},\tilde{b}),
\end{equation}
whereas Eqs. (\ref{L1}) - (\ref{L2}) do not have the corresponding
symmetry
\begin{equation}\label{CC2}
    (a,b,c) \rightarrow (b,c,a)
    \rightarrow (c,a,b).
\end{equation}

The difference results from the fact that Eqs. (\ref{L1}) -
(\ref{L2}) has been obtained by imposition onto the original set
of equations defining the nondiagonal dynamics (see, Eqs.
(2.14)-(2.20) in \cite{BKL2}) the condition
\begin{equation}\label{CC3}
\Gamma_1 >> \Gamma_2 >> \Gamma_3 \, ,
\end{equation}
which implies  Eq. (\ref{AA}).

\subsection{Dynamical system analysis}

Introducing the notation:
\begin{equation}\label{ss1}
x_1 := \ln\tilde{a},~~ x_2 := \ln\tilde{b}, ~~ x_3 := \ln\tilde{c},
~~  p_1 := \dot{x}_1, ~~ p_2 := \dot{x}_2, ~~ p_3 := \dot{x}_3,
\end{equation}
we rewrite the system (\ref{s1}) - (\ref{s3}) as follows

\begin{align}
\dot{x}_1 &= p_1, \label{xx1}\\
\dot{x}_2 &= p_2,\label{xx2}\\
\dot{x}_3 &= p_3,\label{xx3}\\
\dot{p}_1 &=  (e^{x_2}-e^{x_3})^2 - e^{2 x_1} , \label{pp1}\\
\dot{p}_2 &= (e^{x_3}-e^{x_1})^2 - e^{2 x_2} , \label{pp2}\\
\dot{p}_3 &= (e^{x_1}-e^{x_2})^2 - e^{2 x_3} ,\label{pp3}
\end{align}
with the constraint corresponding to (\ref{s4}) in the form
\begin{equation}\label{ss2}
p_1 p_2 + p_1 p_3 + p_2 p_3 - (e^{2 x_1} + e^{2 x_2} + e^{2
x_3})+2 (e^{x_1 + x_2} + e^{x_1 + x_3} + e^{x_2 + x_3}) =0 .
\end{equation}

 It is easy to see that the critical points of the vector field
(\ref{xx1}) - (\ref{pp3}), satisfying (\ref{ss2}), are defined by
\begin{align}
\tilde{S}_{B0}&:= \{(x_1,x_2,x_3,p_1,p_2,p_3)\in \bar{\dR}^6 ~|~ (x_1,x_2,x_3
\rightarrow -\infty)\wedge (p_1 = 0 = p_2 = p_3) \},\label{ss31}\\
\tilde{S}_{B1}&:= \{(x_1,x_2,x_3,p_1,p_2,p_3)\in \bar{\dR}^6 ~|~ (x_1
\rightarrow -\infty,~ x_2 = x_3 )\wedge (p_1 = 0 = p_2 = p_3) \},\label{ss32}\\
\tilde{S}_{B2}&:= \{(x_1,x_2,x_3,p_1,p_2,p_3)\in \bar{\dR}^6 ~|~ (x_2
\rightarrow -\infty,~ x_3 = x_1 )\wedge (p_1 = 0 = p_2 = p_3) \},\label{ss33}\\
\tilde{S}_{B3}&:= \{(x_1,x_2,x_3,p_1,p_2,p_3)\in \bar{\dR}^6 ~|~
(x_3 \rightarrow -\infty,~ x_1= x_2
 )\wedge
(p_1 = 0 = p_2 = p_3) \}.\label{ss34}
\end{align}
There are  no {\it strong} relations among $x_1,~x_2 $ and $x_3$
in each of the above sets, contrary to the nondiagonal case (see
the statement following Eq. (\ref{critS0})).\\

One can solve the constraint equation (\ref{ss2}) setting, e.g.
\begin{equation}
p_3=\frac{e^{2 {x_1}}+e^{2 {x_2}}+e^{2x_3}-2 e^{{x_1}+{x_2}}-2
e^{{x_1}+{x_3}}-2 e^{{x_2}+{x_3}}-{p_1} {p_2}}{{p_1}+{p_2}},
\end{equation}
which turns the vector field (\ref{xx1}) - (\ref{pp3}) into
\begin{eqnarray}
\dot{x}_1 &=& p_1, \label{xx11}\\
\dot{x}_2 &=& p_2,\label{xx22}\\
\dot{x}_3 &=& \frac{e^{2 t{x_1}}+e^{2 {x_2}}+e^{2x_3}-2 e^{{x_1}+{x_2}}-2 e^{{x_1}
+{x_3}}-2 e^{{x_2}+{x_3}}-{p_1} {p_2}}{{p_1}+{p_2}},\label{xx33}\\
\dot{p}_1 &=&  (e^{x_2}-e^{x_3})^2 - e^{2 x_1} , \label{pp11}\\
\dot{p}_2 &=&  (e^{x_3}-e^{x_1})^2 - e^{2 x_2} . \label{pp22}
\end{eqnarray}
The above system has the same critical subspaces as the one
without the constraint built into it. The Jacobian associated with
the system (\ref{xx11}) - (\ref{pp22}) is found to be\\
$$\left(
\begin{array}{ccccc}
 0 & 0 & -\frac{2 e^{{x_1}} \left(-e^{{x_1}}+e^{{x_2}}+e^{{x_3}}\right)}{{p_1}+{p_2}} &
 -2 e^{2 {x_1}} & 2 e^{{x_1}} \left(e^{{x_1}}-e^{{x_3}}\right) \\
 0 & 0 & -\frac{2 e^{{x_2}} \left(e^{{x_1}}-e^{{x_2}}+e^{{x_3}}\right)}{{p_1}+{p_2}} &
 2 e^{{x_2}} \left(e^{{x_2}}-e^{{x_3}}\right) & -2 e^{2 {x_2}} \\
 0 & 0 & -\frac{2 e^{{x_3}} \left(e^{{x_1}}+e^{{x_2}}-e^{{x_3}}\right)}{{p_1}+{p_2}} &
 2 e^{{x_3}} \left(-e^{{x_2}}+e^{{x_3}}\right) & 2 e^{{x_3}} \left(-e^{{x_1}}+e^{{x_3}}\right) \\
 1 & 0 & -\frac{\left(e^{{x_1}}-e^{{x_2}}\right)^2+e^{2 {x_3}}-2 e^{{x_3}}
 \left(e^{{x_1}}+e^{{x_2}}\right)+p_2^2}{({p_1}+{p_2})^2} & 0 & 0 \\
 0 & 1 & -\frac{\left(e^{{x_1}}-e^{{x_2}}\right)^2+e^{2 {x_3}}-2 e^{{x_3}}
 \left(e^{{x_1}}+e^{{x_2}}\right)+p_1^2}{({p_1}+{p_2})^2} & 0 & 0
\end{array}
\right)^T$$ The characteristic polynomial evaluated at the
critical subspaces reads:
\begin{equation}
P(\lambda)=-\lambda^5 .
\end{equation}
Hence, we can conclude that  the character of the  critical
hypersurfaces (\ref{ss31}) - (\ref{ss34}) is the {\it
nonhyperbolic} one.

\subsection{Introducing the qHN variables}

For the diagonal case we define the qHN variables
$(M_\alpha,\Omega_\alpha)$ a follows:
\begin{equation}\label{qq1}
M_1 := \tilde{a}^2 \tilde{V},~~~M_2 := \tilde{b}^2
\tilde{V},~~~M_3 := \tilde{c}^2 \tilde{V},
\end{equation}
\begin{equation}\label{qq2}
\Omega_1 := \tilde{V}\frac{d}{d\tau}\ln\tilde{a} - 1,~~~\Omega_2
:= \tilde{V}\frac{d}{d\tau}\ln\tilde{b} - 1,~~~\Omega_3 :=
\tilde{V}\frac{d}{d\tau}\ln\tilde{c} - 1,
\end{equation}
where $\tilde{V}:= 3/\frac{d}{d\tau} \ln
(\tilde{a}\tilde{b}\tilde{c})$, and $M_\alpha >0, \forall \alpha$,
as near the singularity $\tilde{a}\tilde{b}\tilde{c}\rightarrow
0$.

Making use of (\ref{pp1}) - (\ref{pp3}) we can rewrite the
constraint (\ref{ss2}) in the form:
\begin{equation}\label{qq3}
 p_1 p_2 + p_1 p_3 + p_2 p_3 - (\dot{p}_1 +
\dot{p}_2 + \dot{p}_3)= 0.
\end{equation}
Using (\ref{qq2}) and applying the analysis similar as in the
nondiagonal case (\ref{d1}) - (\ref{sol1}) we get:
\begin{equation}\label{qq4}
p_1 = \Omega (1 + \Omega_1),~~~p_2 = \Omega (1 + \Omega_2),~~~p_3
= \Omega (1 + \Omega_3),
\end{equation}
where $\Omega \in C^1(\dR)$, and where $\Omega_1 + \Omega_2 +
\Omega_3 = 0$ due to (\ref{qq2}).

Now, using (\ref{qq1}) and (\ref{pp1}) - (\ref{pp3}) we arrive to
the expressions:
\begin{eqnarray}
\dot{p}_1 &=& \Omega (-M_1 + M_2 + M_3 - 2 \sqrt{M_2 M_3}), \label{qq5}\\
\dot{p}_2 &=& \Omega (M_1 - M_2 + M_3 - 2 \sqrt{M_1 M_3}), \label{qq6}\\
\dot{p}_3 &=& \Omega (M_1 + M_2 - M_3 - 2 \sqrt{M_1
M_2}).\label{qq7}
\end{eqnarray}
Inserting  (\ref{qq4}) - (\ref{qq7}) into  (\ref{qq3}) leads to
the following expression for the constraint in terms of the qHN
variables:
\begin{equation}\label{qq88}
\Omega\;(\Omega\; \Omega_{123}- M_{123}) = 0,
\end{equation}
where $ \Omega_{123}:= 3 + \Omega_1 \Omega_2 + \Omega_1 \Omega_3 +
\Omega_2 \Omega_3$ and $ M_{123}:= M_1 + M_2 + M_3 - 2(\sqrt {M_1
M_2} + \sqrt {M_1 M_3} + \sqrt {M_2 M_3}).$ Eq. (\ref{qq88}) has
two solutions: $\Omega = 0$, and $\Omega = M_{123}/\Omega_{123}$.

\subsection{The vector field}

Acting with $d/d \tau$ on (\ref{qq1}) and (\ref{qq2}), and using
the expressions (\ref{qq4}) - (\ref{qq7}) leads, after some simple
but lengthy  rearrangements, to the following vector field:

\begin{eqnarray}
\dot{M}_1 &=& \frac{1}{3} M_1 M_{123}(6\Omega_0 (1+\Omega_1) - 1),\label{ap1}\\
\dot{M}_2 &=&  \frac{1}{3}M_2 M_{123}(6 \Omega_0 (1+\Omega_2) - 1) ,\label{ap2}\\
\dot{M}_3 &=&  \frac{1}{3}M_3M_{123} (6 \Omega_0 (1+\Omega_3) - 1),\label{ap3}\\
\dot{\Omega}_1 &=& -M_1 + M_2 + M_3 - 2 \sqrt{M_2 M_3} - \frac{1}{3} M_{123}\; (\Omega_1+1) , \label{ap4}\\
\dot{\Omega}_2 &=& M_1 - M_2 + M_3 - 2 \sqrt{M_1 M_3} -\frac{1}{3} M_{123}\; (\Omega_2 +1), \label{ap5}\\
\dot{\Omega}_3 &=& M_1 + M_2 - M_3 - 2 \sqrt{M_1 M_2} -
\frac{1}{3} M_{123}\; (\Omega_3+1)  ,\label{ap6}
\end{eqnarray}
where  $\Omega_0 = 0$ or $\Omega_0 = 1/\Omega_{123}$, and the
identity  $\Omega_1 + \Omega_2 +\Omega_3 = 0$ must be satisfied.
Equations (\ref{ap4})-(\ref{ap6}) fulfill this identically which
shows self-consistence of the set (\ref{ap1})-(\ref{ap6}).

\subsection{Critical points}

 The critical points of the system (\ref{ap1}) -
(\ref{ap6}), satisfying (\ref{qq88}), define the set of critical
hypersurfaces:
\begin{eqnarray}
{S}_{0}:&=&
\{(\Omega_1,\Omega_2,\Omega_3,M_1,M_2,M_3)\;|\;M_1 = 0 = M_2 =
M_3\}\subset {\bar{\dR}}^6,\label{ac1}\\
{S}_{1}:&=&
\{(\Omega_1,\Omega_2,\Omega_3,M_1,M_2,M_3)\;|\;M_1 = 0,  M_2= M_3\}\subset {\bar{\dR}}^6,\label{ac2}\\
{S}_{2}:&=& \{(\Omega_1,\Omega_2,\Omega_3,M_1,M_2,M_3)\;|\;M_2 =0,
M_3 = M_1
\}\subset {\bar{\dR}}^6,\label{ac3}\\
{S}_{3}:&=& \{(\Omega_1,\Omega_2,\Omega_3,M_1,M_2,M_3)\;|\;M_3=0,
M_1 =  M_2 \}\subset {\bar{\dR}}^6.\label{ac4}
\end{eqnarray}

The Jacobian associated with the vector field (\ref{ap1}) -
\ref{ap6}), satisfying (\ref{qq88}), evaluated at any point of
$\{S_0,S_1,S_2,S_3\}$ has diverging components arising  from
differentiating terms of the type $\sqrt{M_1 M_2}$ and in the
limit $M_1\mapsto0$ (or other $M$'s going to zero). However,
calculating characteristic polynomial and taking the value of its
coefficient at the critical subspaces leads to the following
result:
\begin{equation}\label{jacc11}
P(\lambda)=\lambda^6,
\end{equation}
Hence, we can conclude that the character of the  critical
hypersurfaces (\ref{ac1}) - (\ref{ac4}) is the {\it nonhyperbolic}
one.

\section{Numerical simulations of the dynamics}

In this section we present the numerical simulations of both evolutions, defined by
Eqs. \!\eqref{L1}--\eqref{L2} and \eqref{s1}--\eqref{s4}, to give support to some assumptions
of the preceding sections.
The numerical method we employed here is the same as described in \cite{Nick}.
Our simulations concern the dynamics with  the initial data satisfying the strong
inequality defined by Eq. \!\eqref{CC3}. Since the product of the three scale factors is proportional to the volume
density of the space, decreasing volume means evolution towards the singularity.

\begin{figure}[ht!]
	\centering
	\begin{subfigure}[t]{0.4\textwidth}
		\includegraphics[width=7.0cm,angle=0]{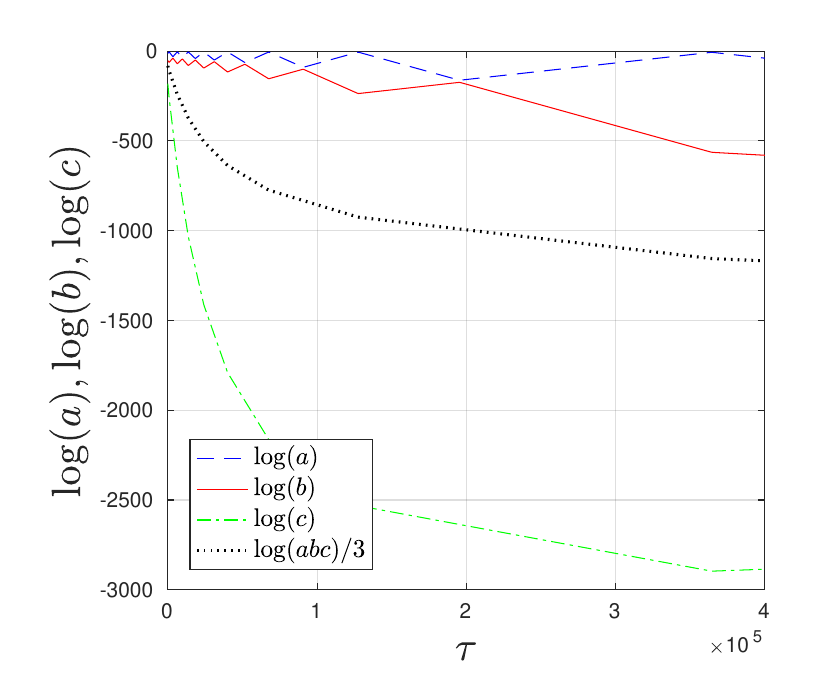}
		\caption{Numerical simulation of the non-diagonal case asymptotically described by the system of equations (\ref{L1}).}
		\label{bkl_nondiagonal}
	\end{subfigure}
	\qquad \quad
	\begin{subfigure}[t]{0.4\textwidth}
		\includegraphics[width=7.0cm,angle=0]{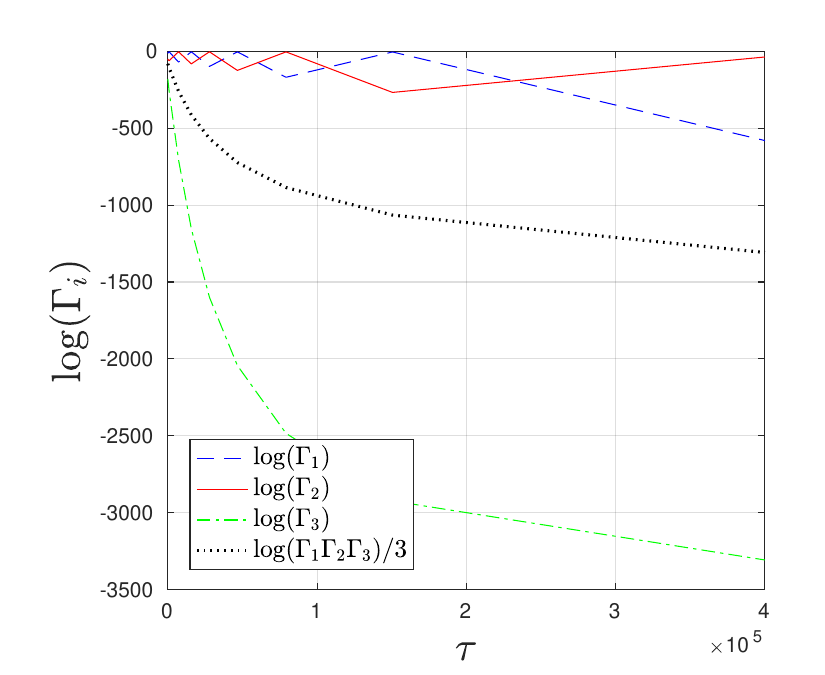}
		\caption{Numerical simulation of  the  diagonal case described by the system of equations (\ref{bb1})-(\ref{bb3}).}
		\label{bkl_diagonal}
	\end{subfigure}
	\caption{Numerical simulations.}
	\label{numerics}
\end{figure}

FIG.~\ref{bkl_nondiagonal} presents the plots of the directional scale factors corresponding to the dynamics of the nondiagonal case.
Taking the initial data satisfying \eqref{CC3} leads to the evolution towards the singularity that {\it maintains}
this strong inequality. This result gives support to the claim that this dynamics has the special {\it asymptotic}
regime.
Further support can be found in \cite{Nick}, where the simulations have been performed by using the exact dynamics of the general 
Bianchi IX model filled with a tilted pressureless fluid.

FIG.~\ref{bkl_diagonal}  presents the evolution of the directional scale factors of the diagonal case with almost the same initial
data as in the nondiagonal case\footnote{The initial data cannot be exactly the same as they must satisfy the
dynamical constraints defined by \eqref{L2} and \eqref{s4} which are different.}. No special regime occurs in this case.
One can see the permutation symmetry of the relation \eqref{CC1} during the evolution of the system, {\it contrary} to the nondiagonal
case. The permutation of the initial data leads to the same solutions (recoloring the plots), which is consistent with
the permutation symmetry of the dynamics \eqref{s1}--\eqref{s4}.
In fact, the permutation symmetry \eqref{CC1} was used to check  the correctness of the numerical simulations.

We were able to keep the numerical error in solving the Hamiltonian constraints,
 \eqref{L2} or \eqref{s4}, as low as  the order of $10^{-16}$. This is illustrated in FIG. 2.
Further increase of the precision of calculations keeps the plots unchanged.

\begin{figure}[ht!]
	\centering
	\begin{subfigure}[t]{0.4\textwidth}
		\includegraphics[width=7.0cm,angle=0]{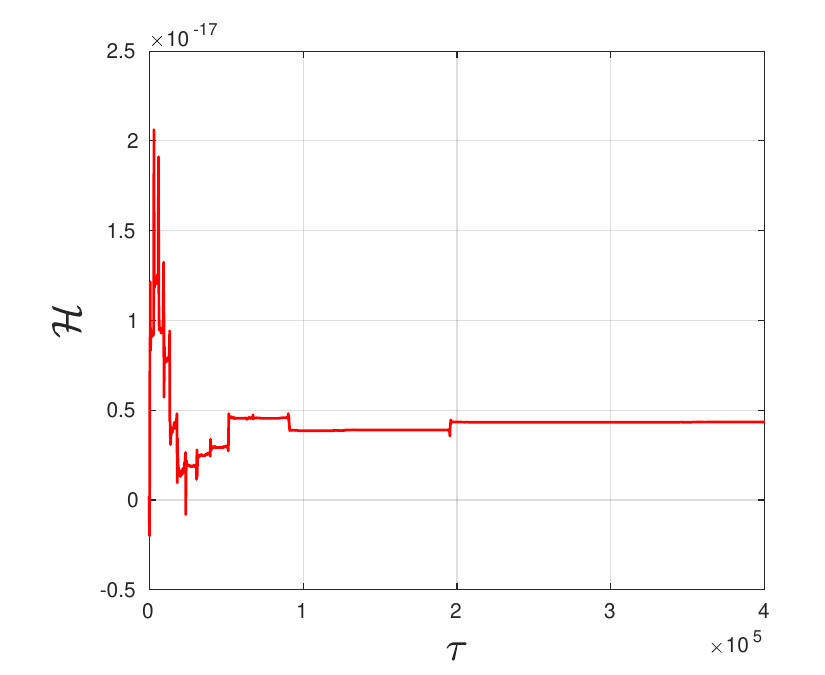}
		\caption{Error in the Hamiltonian constraint
		belonging to FIG.~\ref{bkl_nondiagonal}.}
		\label{error_diagonal}
	\end{subfigure}
	\qquad \quad
	\begin{subfigure}[t]{0.4\textwidth}
		\includegraphics[width=7.0cm,angle=0]{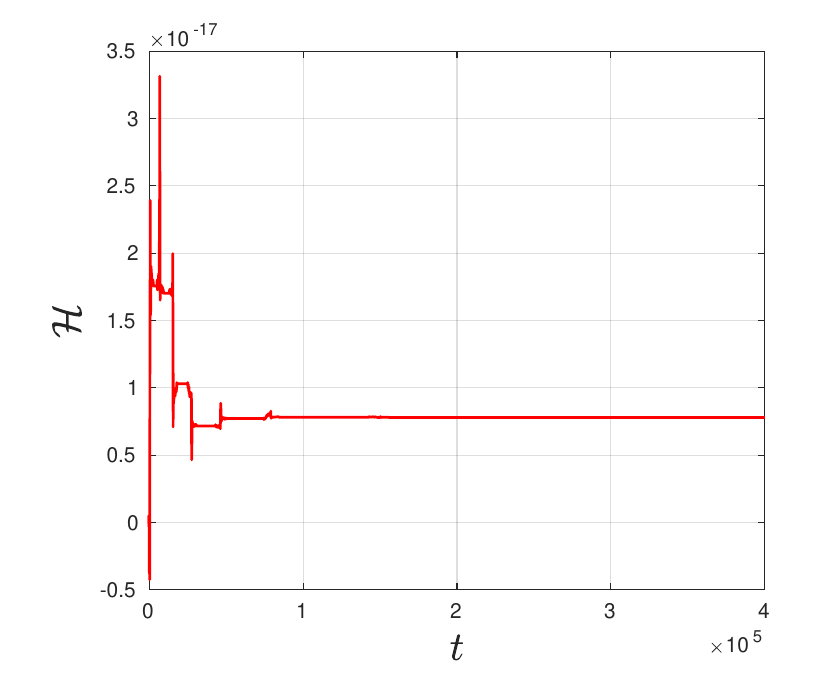}
		\caption{Error in the Hamiltonian constraint
		belonging to FIG.~\ref{bkl_diagonal}.}
		\label{error_non-diagonal}
	\end{subfigure}
	\caption{Error in the numerical simulations.}
	\label{numerics_error}
\end{figure}


\section{Conclusions}

Near the cosmological singularity, an evolution of the  Bianchi IX model is
an infinite sequence of the so called eras each of which consists of
the Kasner type epochs \cite{BKL22}.  In the {\it diagonal} case, each epoch
can be described, e.g.,  by the
relation $\tilde{\Gamma}_1 \sim \tilde{\Gamma}_2 >
\tilde{\Gamma}_3$ (where $\sim$ means coupled) called an
oscillation\footnote{There can also occur small oscillations
$\tilde{\Gamma}_1 \sim \tilde{\Gamma}_2 >> \tilde{\Gamma}_3$, but
they last for a finite interval of time and can be ignored.}. The
dynamics of the {\it nondiagonal} model has  essentially different
structure \cite{Belinski:2014kba,BKL2}: the oscillation of the
diagonal type, e.g., $\Gamma_1
\sim \Gamma_2 > \Gamma_3$ enters sooner or later the relation
$\Gamma_1 > \Gamma_2 > \Gamma_3$, which turns into the strong
relation $\Gamma_1 >> \Gamma_2 >> \Gamma_3$. Finally, the
system approaches the singularity in a finite proper time.

The difference between the dynamics of the diagonal and
nondiagonal cases leads to different topological structures of the
corresponding sets of critical points. In the former case, this
set consists of three hypersurfaces in ${\bar{\dR}}^6$  having the
same  topology, Eqs. \eqref{ac2}-\eqref{ac4},   and one set, Eq.
\eqref{ac1}, with the simple topology of ${\bar{\dR}}^3$. In the
latter case, the set of critical points has sophisticated
topology,  defined by Eq. \eqref{critnew}, quite different from
the diagonal case. Similar relationship occurs between the critical sets
expressed in term of the BKL variables. However, in both cases the
critical sets consist of the {\it nonhyperbolic} type of critical
points.

The nonhyberbolicity is expected to be directly linked with the chaoticity
of the dynamics of the Bianchi IX model.
We conjecture that due to the different topologies of the critical spaces the chaoticity
aspects of both cases can be different.  Further studies are required to get
insight into this intriguing issue.

Our main concern is the nonhyperbolicity of equilibrium points in both
diagonal and general cases. They do not define a set of {\it isolated}
points, but a three-dimensional {\it continuous} space.  Thus,  our
choice of phase space variables seems to be unsatisfactory. We have
already tried \cite{Czuchry:2012ad} to use the so-called blowing up
technic initiated by McGehee \cite{McG} to avoid this obstacle,
but with no success. More sophisticated approach based on
$\sigma$-process of algebraic geometry proposed in \cite{Bogo1} may
bring some progress, but it leads to a noncanonical variables that
we try to avoid.  Another framework proposed for the spacially
inhomogeneous models \cite{Uggla:2003fp}, within Hubble-normalized
approach, can be probably specialized to the homogeneous models.
However, this formulation is again a noncanonical one which we do
not favour.

The way out seems to be giving up the insistence on dealing entirely
with canonical formulations and planning making use of coherent states
quantization methods (based on phase space structure of the
underlying system)  that we have recently applied to the diagonal
Bianchi IX model \cite{Bergeron:2015lka,Bergeron:2015ppa}. In such a case making
use of the results of \cite{Uggla:2003fp} to elucidate
mathematical structure of the physical phase space specific to the
dynamics of the Bianchi IX model (in both considered cases) would
make sense. This is supposed to be the next step of our
investigation and the results of the present paper could be used
as a starting point. Another approach would be based on modification
of the definition of the Hubble-normalized variables that we use in the
present paper.

The fact that some critical points occur at infinity is not an
obstacle. The mapping of the set of critical points onto the
Poincar\'{e} sphere (considered, e.g.,  for the nondiagonal case,
in App. C) des not change the type of the criticality. It stays
to be of nonhyperbolic type.
Thus, compactification of phase space does not  help.

It seems  that the nonhyperbolicity of the equilibrium
points distributed in a continuous way in higher dimensional space
is a generic feature  of the dynamics of the Bianchi IX
model and cannot be avoided.  These properties  may correspond to
mathematical structure \cite{WE,Aul} underlying chaotic behaviour
of considered dynamics (see, e.g., \cite{JB1,JB2}), and needs to be further examined.

\acknowledgments We are grateful to Claes Uggla for the suggestion
to use the Hubble normalized variables and to Juliette Hell for
valuable discussions concerning the dynamics of the Bianchi IX
model.  Finally, we appreciate inspiring discussions with Vladimir
Belinski. This work was partially supported by the German-Polish bilateral project DAAD and
MNiSW, No 57391638, ``Model of stellar collapse towards a singularity and
its quantization''

\appendix

\section{Quotient coordinates}

In order to avoid defining critical surface in term of the limits
$\sqrt{N_3/N_2} << \sqrt{N_2/N_1} << N_1^2 \rightarrow 0$, one can
introduce quotient coordinates:
\begin{eqnarray}
u &:=& \frac{1}{N_1^2}\sqrt{\frac{N_2}{N_1}},\label{udef}\\
v &:=& \frac{1}{N_1^2}\sqrt{\frac{N_3}{N_2}}.\label{vdef}
\end{eqnarray}
Then the system of equations  (\ref{fin1}) - (\ref{fin5})  takes
the following form:
\begin{eqnarray}\dot{N_1}&=&-\frac{{N_1}^2}{3}-\frac{N_1^2(1+{\Sigma_1})  \left(1+
\sqrt{1-4 (u+
v)\Sigma}\right)}{\Sigma},\label{quot1}\\
\dot{u}&=&u{N_1}  \frac{\Sigma_0+3(4+5 {\Sigma_1}- {\Sigma_2} )\sqrt{1-4
(u+v) \Sigma}}{6 \Sigma },\label{quot2}\\
\dot{v}&=&v{N_1}  \frac{\Sigma_0+3(4+ 5{\Sigma_1}+2 {\Sigma_2})\sqrt{1-4
(u+v) \Sigma}}{6 \Sigma},\label{quot3}\\
\dot{\Sigma_1}&=&-\frac{1}{3} (4+{\Sigma_1}) {N_1}-\frac{2
u
{N_1}\Sigma}{1+\sqrt{1-4  (u+v)\Sigma}},\label{quot4}\\
\dot{\Sigma_2}&=&-\frac{1}{3} (-2+{\Sigma_2}) {N_1}+\frac{2
(u-v)
{N_1}\Sigma}{1+\sqrt{1-4  (u+v)\Sigma}}\label{quot5}
\end{eqnarray}
where $\Sigma_0=15 {\Sigma_1}+4 {\Sigma_1}^2-3 {\Sigma_2}+4
{\Sigma_1} {\Sigma_2}+ 4 {\Sigma_2}^2$ and $\Sigma= -3+{\Sigma_1}^2+
{\Sigma_1} {\Sigma_2}+{\Sigma_2}^2$. The left hand sides of
equations (\ref{quot1})-(\ref{quot2}) vanish for $N_1=0=u=v$.

 The set of critical points of the vector field
(\ref{quot1})-(\ref{quot5}) is easily found to be
\begin{equation}\label{quocoor}
\tilde{S}_{qHN}:= \{(\Sigma_1,\Sigma_2,N_1,u,v)\;|\;N_1 = 0 = u =
v\}\subset {\bar{\dR}}^5 \, .
\end{equation}
The characteristic polynomial is  $P(\lambda)=-\lambda^5$. Thus,
the character of corresponding critical surface is nonhyperbolic.

One may speculate that $\tilde{S}_{qHN}$ corresponds to $S_0$ of
Eq. (\ref{ac1}) so the underlying dynamics of corresponding vector
fields have some common feature. One may further speculate that
both $S_0$ of Eq. (\ref{ac1}) and \eqref{quocoor} correspond to
some new form of chaoticity, whereas  \eqref{ac2}-\eqref{ac4} are
specific to the well known attractor of the diagonal case.

\section{Relationship between old and  new  variables}

Let us rewrite Eqs. (\ref{L1})  as follows

\begin{eqnarray}
\dot{q}_1 &=& \pi_1, \label{x1}\\
\dot{q}_2 &=& \pi_2,\label{x2}\\
\dot{q}_3 &=& \pi_3,\label{x3}\\
\dot{\pi}_1 &=& - \exp(2 q_1) + \exp (q_2 -q_1), \label{p1}\\
\dot{\pi}_2 &=& \exp(2 q_1) - \exp (q_2 - q_1)+ \exp (q_3 - q_2), \label{p2}\\
\dot{\pi}_3 &=& \exp(2 q_1) - \exp (q_3 -q_2),\label{p3}
\end{eqnarray}
where $q_1 := \ln a,\; q_2 := \ln b,\; q_3 := \ln c, \; \pi_1 :=
\dot{q}_1,\;\pi_2 := \dot{q}_2,\;\pi_3 := \dot{q}_3 $ are new
variables. Thus, the constraint (\ref{L2}) reads
\begin{equation}\label{a3}
\pi_1 \pi_2 + \pi_1 \pi_3 + \pi_2 \pi_3 = \exp(2 q_1) + \exp (q_2
- q_1)+ \exp (q_3 - q_2) .
\end{equation}
Making use of (\ref{x1})-(\ref{p3}) one can present (\ref{a3}) in
the form
\begin{equation}\label{aa3}
\pi_1 \pi_2 + \pi_1 \pi_3 + \pi_2 \pi_3  = 4 \dot{\pi}_1 + 3
\dot{\pi}_2 + 2 \dot{\pi}_3 .
\end{equation}

One can easily verify that the critical points of the dynamical
system (\ref{x1})-(\ref{a3}) are of the {\it nonhyperbolic} type
and coincide with the set of critical points $S_B$ determined in
\cite{Czuchry:2012ad}. Thus,  the set  of critical points $S_B$
(in terms of $q_\alpha$ and $\pi_\alpha$ variables) is given by
\bea \label{critS}S_B: &=& \{(q_1,q_2,q_3,\pi_1,\pi_2,\pi_3)\in
\bar{\dR}^6 ~|~ (q_1 \rightarrow
    -\infty,~ q_2-q_1 \rightarrow -\infty,~ q_3-q_2 \rightarrow
    -\infty)\nonumber
    \\&& \wedge (\pi_1 = 0 = \pi_2 = \pi_3  \label{critS0}\}, \eea
where $\bar{\dR}:= \dR \cup  \{-\infty, +\infty\}$.  The
infinities in (\ref{critS0}) should be approached in such a way
that $q_1 \gg q_2 \gg q_3$, which corresponds to $a \gg b \gg c $
found in \cite{BKL2}.

 Now, we rewrite the vector field (\ref{x1})-(\ref{a3}) in terms
of the qHN variables $N_\alpha$ and $\Sigma_\alpha$.  Using
(\ref{a2}) we get
\begin{equation}\label{d1}
\Sigma_\alpha = 3\pi_\alpha /(\pi_1 + \pi_2 + \pi_3) - 1.
\end{equation}
Eq. (\ref{d1}) can be presented in a matrix form as follows

\begin{equation}\label{mat1}
 \left[\begin{array}{ccc}
  \Sigma_1 - 2& 1+ \Sigma_1 & 1+\Sigma_1  \\
  1+\Sigma_2 & \Sigma_2 -2 & 1+ \Sigma_2  \\
  1+\Sigma_3 & 1+\Sigma_3 & \Sigma_3 -2  \\
\end{array}\right]
\left[\begin{array}{c} \pi_1 \\ \pi_2 \\ \pi_3 \end{array}\right]
= \left[\begin{array}{c} 0 \\ 0 \\ 0  \end{array}\right].
\end{equation}
One may verify that the determinant of the 3 x 3 matrix $A$ of the
above equation reads: $det(A) = 9 \,(\Sigma_1 + \Sigma_2 +
\Sigma_3) = 0$, since $\Sigma_1 + \Sigma_2 + \Sigma_3 =0$. Thus,
rank of $A <3$. One may easily check that all minors
($M_k,~k=1,2,3$) of the $2 \times$ 2 submatrixes of the  matrix
$A$ are of the form $M_k =\pm 3 (1+\Sigma_k)$. Since we cannot
have $1+\Sigma_k =0, \forall k $ due to $\Sigma_1 + \Sigma_2 +
\Sigma_3 =0$, the rank of the $A$ matrix equals $2$. Suppose we
choose
\begin{equation}\label{mat2}
B= \left[\begin{array}{cc}
  \Sigma_1 - 2& 1+ \Sigma_1  \\
  1+\Sigma_2 & \Sigma_2 -2   \\
\end{array}\right]
\end{equation}
to play the role of a nonsingular submatrix of $A$. Since $det B =
3(1+ \Sigma_3)$, the rank of $B$ equals $2$ if we have
\begin{equation}\label{con55}
1+ \Sigma_3 \neq 0.
\end{equation}

Using Cramer's rules we find the following solution to
(\ref{mat1}):
\begin{equation}\label{sol1}
\pi_1 = \Pi (1+\Sigma_1),~~~\pi_2 = \Pi (1+\Sigma_2)~~~\pi_3 = \Pi
(1+\Sigma_3),
\end{equation}
where  we have redefined an arbitrary variable $ \pi_3 = \Pi \in
C^1(\dR)$ by taking $\pi_3=\Pi (1+\Sigma_3)$, which is allowed as
$(1+\Sigma_3)\neq0$. It is clear that one can  get the solution
(\ref{sol1}) assuming that either $1+ \Sigma_1 \neq 0$ or $1+
\Sigma_2 \neq 0$, instead of (\ref{con55}). Therefore, our
solution (\ref{sol1})  is independent on the choice of the minor
$M_k$ connected with the matrix $A$ of (\ref{mat1}). We conclude
that the general solution to the matrix equation (\ref{mat1}) is
defined by (\ref{sol1}).

Using (\ref{a1})  we obtain
\begin{equation}\label{d2}
N_\alpha = 3\exp(2q_\alpha)/(\pi_1 +\pi_2 + \pi_3),
\end{equation}
that   leads to
\begin{equation}\label{b2}
\exp (2q_1) = N_1 (\pi_1 +\pi_2 + \pi_3)/3,~~~\exp (q_2 - q_1)=
\sqrt{N_2/N_1},~~~\exp (q_3 - q_2)= \sqrt{N_3/N_2}.
\end{equation}

Combining (\ref{p1}) - (\ref{p3})  we obtain
\begin{equation}\label{b3}
\exp(2q_1) = \dot{\pi}_1 + \dot{\pi}_2 + \dot{\pi}_3,~~~\exp (q_2
- q_1)= 2 \dot{\pi}_1 + \dot{\pi}_2 + \dot{\pi}_3,~~~\exp (q_3 -
q_2)= \dot{\pi}_1 + \dot{\pi}_2.
\end{equation}
Comparing (\ref{b2}) with (\ref{b3}), and using the solution
(\ref{sol1}), we  get
\begin{eqnarray}
 \dot{\pi}_1 + \dot{\pi}_2 + \dot{\pi}_3&=& \Pi N_1, \label{y1}\\
 2 \dot{\pi}_1 + \dot{\pi}_2 + \dot{\pi}_3 &=& \sqrt{N_2/N_1},\label{y2}\\
 \dot{\pi}_1 + \dot{\pi}_2 &=& \sqrt{N_3/N_2},\label{y3}
\end{eqnarray}
Which can be presented in a matrix form as follows:
\begin{equation}\label{mat3}
\left[\begin{array}{ccc}
  1& 1 & 1  \\
  2 & 1 & 1  \\
  1 & 1 & 0  \\
\end{array}\right]
\left[\begin{array}{c} \dot{\pi}_1 \\ \dot{\pi}_2 \\ \dot{\pi}_3
\end{array}\right] = \left[\begin{array}{c} \Pi N_1 \\ \sqrt{N_2/N_1} \\ \sqrt{N_3/N_2}
\end{array}\right].
\end{equation}
One may easily verify that determinant of the matrix defining
(\ref{mat3}) equals one, so the system has only one solution. It
is found to be:
\begin{equation}\label{sol2}
\dot{\pi}_1 = \sqrt{N_2/N_1} - \tilde{\Pi} ,~~~\dot{\pi}_2 =
\sqrt{N_3/N_2} - \sqrt{N_2/N_1} + \tilde{\Pi},~~~\dot{\pi}_3 =
-\sqrt{N_3/N_2} + \tilde{\Pi},
\end{equation}
where $\tilde{\Pi} :=\Pi N_1$.

An arbitrary variable $\Pi $ that occurs in (\ref{sol1}) and
(\ref{sol2}) can be fixed by the constraint (\ref{aa3}). It leads
to the following equation for $\Pi$:
\begin{equation}\label{conPi}
{(3+ \Sigma_1 \Sigma_2 + \Sigma_1 \Sigma_3+ \Sigma_2 \Sigma_3)}
\Pi^2 - { N_1}\, \Pi -  \big(\sqrt{N_2/N_1} +  \sqrt{N_3/N_2}\big)
= 0,
\end{equation}
where $\Sigma_1 + \Sigma_2 + \Sigma_3 = 0$.

\section{The Poincar\'{e} type variables}


Since examination  of phase space at `infinite region',
(\ref{critS}), is difficult mathematically, we change coordinates
of the phase space to map the set of critical points (\ref{critS})
onto a finite region. We map the infinite space  $\bar{\dR}^6$
into a finite Poincar\'{e} sphere, parameterized by  Cartesian
coordinates $(X_1,X_2,X_3,P_1,P_2,P_3)$, as follows:
 \bea
x_1 &=:& \frac{X_1}{1-r} , \label{poin111} \\
x_2 &=:& \frac{X_2}{1-r},  \label{poin222} \\
x_3 &=:& \frac{X_3}{1-r},  \label{poin223} \\
p_1 &=:& \frac{P_1}{1-r} , \label{poin333}\\
p_2 &=:& \frac{P_2}{1-r} , \label{poin444}\\
p_3 &=:& \frac{P_3}{1-r},\label{poin555} \eea where $r^2=X_1^2 +
X_2^2 +X_3^2+ P_1^2 + P_2^2 + P_3^2 $, and where we redefined the
variables: $x_k := q_k, p_k := \pi_k ~~(k = 1,2,3) $ to get the
connection with the results of our previous paper
(see, Eq. (38) of \cite{Czuchry:2012ad}). We also rescale the time parameter $\tau$
by defining the new time parameter $T$ such that $d{T} :=
d\tau/(1-r)$. In these coordinates our phase space is contained
within a sphere of radius one -- `infinities' correspond to $r=1$.

If the mapping is {\it canonical}, we should have:
\begin{equation}\label{xp2}
\{X_l,X_k\}_{x,p} = 0 = \{P_l,P_k\}_{x,p},~~~~ \{X_l,P_k\}_{x,p} =
\delta_{lk}.
\end{equation}
The map (\ref{poin111})-(\ref{poin555}) is not canonical, because
we have:
\begin{equation}\label{xx}
\{X_k,X_l\}_{x,p} = (1-r) g(a) (x_k p_l - x_l p_k),
\end{equation}
\begin{equation}\label{pp}
\{P_k,P_l\}_{x,p} = (1-r) f(a) (x_k p_l - x_l p_k),
\end{equation}
\begin{equation}\label{xp}
\{X_k,P_l\}_{x,p} = (1-r)^2 \delta_{kl} - (1-r) \big(f(a) x_k x_l
+ g(a) p_k p_l\big),
\end{equation}
where $a:= r^2/ (1-r)^2,~f(a)\neq 0,~g(a)\neq 0 $. It is clear
that there is no chance to get (\ref{xp2}) for any $r$
including the limit $r\rightarrow 1$.

The insertion of (\ref{poin111})-(\ref{poin555}) into
(\ref{x1})-(\ref{p3}) gives:
\begin{eqnarray}
\Big(\frac{X_1}{1-r}\Big)^\prime &=& \frac{1}{2}(-P_1 +P_2 +P_3), \label{6d1}\\
\Big(\frac{X_2}{1-r}\Big)^\prime &=& \frac{1}{2}(P_1 -P_2
+P_3),\label{6d2}\\
\Big(\frac{X_3}{1-r}\Big)^\prime &=& \frac{1}{2}(P_1 +P_2 -P_3), \label{6d3}\\
\Big(\frac{P_1}{1-r}\Big)^\prime &=&
(1-r)\big(2\exp{\frac{2X_1}{1-r}}-
\exp{\frac{X_2 -X_1}{1-r}}\big), \label{6d4}\\
\Big(\frac{P_2}{1-r}\Big)^\prime &=& (1-r)\big(\exp{\frac{X_2 -
X_1}{1-r}}-
\exp{\frac{X_3 -X_2}{1-r}}\big), \label{6d5}\\
\Big(\frac{P_3}{1-r}\Big)^\prime &=& (1-r)\big(\exp{\frac{X_3
-X_2}{1-r}}\big), \label{6d6}
\end{eqnarray}
where prime denotes derivative with respect to the new time
parameter $T$.

To find the fixed points  we insert  ${X}_1' =0={X}_2' = {X}_3' =
{P}_1' ={P}_2' = {P}_3'$ into (\ref{6d1})-(\ref{6d6}) by using the
elementary formulas:
\begin{equation}\label{r6a}
    {r}'= \frac{d}{dT}r = \big(X_1{X}_1'
+X_2 {X}_2' + X_3 {X}_3' + P_1 {P}_1'+ P_2 {P}_2' + P_3
{P}_3'\big)/r
\end{equation}
and, e.g.
\begin{equation}\label{r6b}
\frac{d}{dT}\Big(\frac{X_1}{1-r}\Big) = \frac{{X}_1' (1-r)+ X_1
{r}'}{(1-r)^2}.
\end{equation}
After rearrangement of terms we finally get:
\begin{eqnarray}
-P_1 + P_2 +P_3 &=& 0, \label{6f1}\\
P_1 - P_2 +P_3 &=& 0, \label{6f2}\\
P_1 + P_2 -P_3 &=& 0, \label{6f3}\\
2\exp{\frac{2X_1}{1-r}}- \exp{\frac{X_2 -X_1}{1-r}} &=& 0, \label{6f4}\\
\exp{\frac{X_2 - X_1}{1-r}}- \exp{\frac{X_3 -X_2}{1-r}}&=& 0, \label{6f5}\\
\exp{\frac{X_3 -X_2}{1-r}}&=& 0. \label{6f6}
\end{eqnarray}
The solution to (\ref{6f1})-(\ref{6f3}) reads: $P_1 = 0 = P_2 =
P_3$. The equations (\ref{6f4})-(\ref{6f6}) can  be satisfied in
the limit $r\mapsto1$ if
\begin{equation}\label{6dcond}
\lim_{r\rightarrow1^-}\exp\frac{2X_1}{1-r}= 0 =
\lim_{r\rightarrow1^-}\exp\frac{X_2-X_1}{1-r} =
\lim_{r\rightarrow1^-}\exp\frac{X_3-X_2}{1-r},
\end{equation}
which leads to the condition: $X_3 < X_2 < X_1 < 0.$ Therefore,
the critical subspace is defined to be:
\begin{equation}\label{6S}
S_P:= \{(X_1,X_2,X_3, P_1,P_2,P_3)~|~(X_3 < X_2 < X_1 < 0)\wedge
(P_1 = 0 = P_2 = P_3)\}.
\end{equation}
It is not difficult to verify that the transformation
(\ref{poin111})-(\ref{poin555}) does not map $S_B$ into $S_P$.

It is clear that any point of $S_P$, in the limit $r\rightarrow
1^-$, satisfies the constraint (\ref{a3}) which in the variables
(\ref{poin111})-(\ref{poin555}) has the form:
\begin{eqnarray}\label{con6P}
\frac{1}{2(1-r)^2}(P_1 P_2 + P_1 P_3 + P_2 P_3) -
\frac{1}{4(1-r)^2}(P_1^2 + P_2^2 + P_3^2 ) \\-
\exp\frac{2X_1}{1-r} - \exp\frac{X_2-X_1}{1-r} -
\exp\frac{X_3-X_2}{1-r} = 0.
\end{eqnarray}

One can resolve (either manually or by symbolic computations) the
nonlinear vector field (\ref{6d1})-(\ref{6d6}) with respect to the
derivatives $\;X_1', X_2',\ldots, P_3'$, and find the
corresponding Jacobian. Its value at any point of the subspace
$S_P$ (in the limit $r\mapsto1$) turns out to be a six dimensional
zero matrix. It means that linearization of the exact vector
field, at the set of critical points $S_P$, cannot help in the
understanding of the mathematical structure of the space of orbits
of considered vector field. An examination of the nonlinearity
cannot be avoided. One may say, formally, that the set $S_P$
consists of the {\it nonhyperbolic} type of fixed points.

\end{document}